\documentclass[prd,aps,eqsecnum,twoside]{revtex4-2}

\usepackage{hyperref,amsfonts,amsmath,amssymb,graphicx,bm,subfigure}

\begin{document}

\author{Maxim Dvornikov}
\email{maxdvo@izmiran.ru}

\title{Interaction of supernova neutrinos with stochastic gravitational waves}

\affiliation{Pushkov Institute of Terrestrial Magnetism, Ionosphere
and Radiowave Propagation (IZMIRAN),
108840 Moscow, Troitsk, Russia}


\providecommand{\tabularnewline}{\\}



\begin{abstract}
We examine the propagation and flavor oscillations of neutrinos under
the influence of gravitational waves (GWs) with an arbitrary polarization.
We rederive the effective Hamiltonian for the system of three neutrino
flavors using the perturbative approach. Then, using this result,
we consider the evolution of neutrino flavors in stochastic GWs with
a general energy density spectrum. The equation for the density matrix
is obtained and solved analytically in the case of three neutrino flavors. As an application,
we study the evolution of the flavor content of a neutrino beam emitted in a core-collapsing supernova. We obtain the analytical expressions for the contributions
of GWs to the neutrino fluxes and for the damping decrement, which
describes the attenuation of the fluxes to their asymptotic values.
We find that the contribution to the evolution of neutrino fluxes from GWs, emitted by merging supermassive black
holes, dominates over that from black holes with stellar masses.
The implication of the obtained results for the measurement
of astrophysical neutrinos with neutrino telescopes is discussed.
\end{abstract}

\maketitle

\section{Introduction}

Neutrinos provide a unique possibility to explore physics beyond the
standard model with the help of nonaccelerator methods. Such achievements of neutrino physics became possible
after the observation of oscillations of atmospheric and solar neutrinos~\cite{Nobel2015}.
These experimental facts are the straightforward indications of the
nonzero masses and mixing between different neutrino flavors. External
fields, e.g., the neutrino interaction with background matter~\cite{Smi05}
and electromagnetic fields~\cite{Giu19}, are known to modify the
process of neutrino oscillations.

The gravitational interaction, in spite of its weakness, was found
in Refs.~\cite{PirRoyWud96,CarFul97,For97} to contribute to the neutrino
oscillations dynamics. In the majority of studies, neutrino oscillations
in curved spacetime were examined when particles move in static gravitational
backgrounds, e.g., in the vicinity of a black hole (BH). It is interesting
to analyze the propagation and oscillations of astrophysical neutrinos
in time dependent metrics, e.g., induced by a gravitational wave (GW).

The studies of the fermions evolution in GWs were carried out in Refs.~\cite{Qua16,ObuSilTer17}.
Neutrino spin oscillations, i.e. when we deal with transitions between
active left polarized and sterile right polarized particles, in background
matter under the influence of GW were discussed in Ref.~\cite{Dvo19a}.
Neutrino flavor oscillations in GWs, as well as in gravitational fields
caused by metric perturbations in the early universe, were considered
in Ref.~\cite{KouMet19}.

In this paper, we continue the research in Refs.~\cite{Dvo19,Dvo20},
where the influence of stochastic GWs on neutrino
flavor oscillations was considered. The main problem of Refs.~\cite{Dvo19,Dvo20}
was the consideration of astrophysical neutrinos emitted in decays
of charged pions. Although such neutrinos form the major cosmic neutrinos
background, their sources are distributed rather uniformly in the
universe. Thus, the integral flux in a terrestrial detector should
be averaged over the propagation distance of such neutrinos. This
fact makes it difficult to separate the contribution of stochastic
GWs on the measured flavor composition. To avoid this difficulty, we decide to examine
the effect of stochastic GWs on supernova (SN) neutrinos. If an explosion of a core-collapsing SN
happens in our galaxy, firstly, it emits a sizable neutrino
flux to be measured even by existing neutrino telescopes~\cite{Sch18}.
Secondly, SN is almost a point-like neutrino source. Hence one should
not average over the neutrino propagation distance. In this situation, we expect that 
the effect of stochastic GWs is not smeared.

The present work is motivated by the recent direct detection of GWs
by the LIGO-Virgo collaborations~\cite{Abb16}. There are active
multimessenger searches of GWs and high energy neutrinos by existing
detectors~\cite{Alb19,Aar20a} and suggestions to implement them
in future ones~\cite{Aar20b}. There are also attempts to observe stochastic
GWs~\cite{Per19,Arz20}, with various methods for their detection
being developed~\cite{RomCor17}.

SN neutrinos were reliably detected in 1987 after the SN explosion
in the Large Magellanic Cloud (see, e.g., Ref.~\cite{Raf96}). Since
then, the experimental techniques in construction of neutrino telescopes
made great achievements. Now, if a nearby SN in our galaxy explodes,
a huge number of events will be recorded~\cite{VitTamRaf20}. As
mentioned above, a simultaneous detection of GWs and SN neutrinos
may be possible. Besides a direct neutrino signal from a certain SN,
all collapsing stars in the universe emit neutrinos which form the
diffuse SN neutrino background. There are prospects to measure it
by existing and future neutrino telescopes (see, e.g., Ref.~\cite{An16}).

This work is organized in the following way. In Sec.~\ref{sec:DENSMATR},
we formulate the problem of the propagation of flavor neutrinos in
a plane GW with an arbitrary polarization. Then, we derive the equation
for the density matrix for flavor neutrinos if we deal with stochastic
GWs. This equation is exactly solved for the arbitrary energy spectrum
of GWs. Then, in Sec.~\ref{sec:APPL}, we apply our results for the
description of the interaction of SN neutrinos with stochastic GWs.
We find the corrections to neutrino fluxes and the damping decrement
in an explicit form. Finally, we summarize our results in Sec.~\ref{sec:CONCL}.
The effective Hamiltonian for flavor oscillations under the influence
of GW is rederived in Appendix~\ref{sec:DERHG}.

\section{Evolution of flavor neutrinos in the GWs background\label{sec:DENSMATR}}

The system of three active flavor neutrinos $\nu_{\lambda}$, $\lambda=e,\mu,\tau$,
with the nonzero mixing, as well as under the influence of a plane GW
with an arbitrary polarization, obeys the following Schr\"odinger equation:
\begin{equation}\label{eq:Schreq}
  \mathrm{i}\dot{\nu}=(H_{0}+H_{1})\nu,
\end{equation}
where $\nu^\mathrm{T} = (\nu_e,\nu_\mu,\nu_\tau)$, $H_{0}=UH_{m}^{(\mathrm{vac})}U^{\dagger}$ is the effective
Hamiltonian for vacuum oscillations in the flavor eigenstates basis,
$H_{m}^{(\mathrm{vac})}=\tfrac{1}{2E}\text{diag}\left(0,\Delta m_{21}^{2},\Delta m_{31}^{2}\right)$
is the vacuum effective Hamiltonian for the mass eigenstates $\psi_{a}$,
$a=1,2,3$, $\Delta m_{ab}^{2}=m_{a}^{2}-m_{b}^{2}$ is the difference
of the squares of masses $m_{a}$ of mass eigenstates, $E$ is the
mean energy of a neutrino beam, and $U$ is the unitary matrix which
relates flavor and mass bases: $\nu_{\lambda}=U_{\lambda a}\psi_{a}$.
To derive $H_{m}^{(\mathrm{vac})}$ in Eq.~(\ref{eq:Schreq}) we
assume that neutrinos are ultrarelativistic and subtract a proper
diagonal term.

The mixing matrix $U$ can be present in the form,
\begin{equation}\label{eq:U3f}
  U=
  \left(
    \begin{array}{ccc}
      1 & 0 & 0\\
      0 & c_{23} & s_{23}\\
      0 & -s_{23} & c_{23}
    \end{array}
  \right)
  \cdot
  \left(
    \begin{array}{ccc}
      c_{13} & 0 & s_{13}e^{-\mathrm{i}\delta_{\mathrm{CP}}}\\
      0 & 1 & 0\\
      -s_{13}e^{\mathrm{i}\delta_{\mathrm{CP}}} & 0 & c_{13}
    \end{array}
  \right)
  \cdot
  \left(
    \begin{array}{ccc}
      c_{12} & s_{12} & 0\\
      -s_{12} & c_{12} & 0\\
      0 & 0 & 1
    \end{array}
  \right),
\end{equation}
where $c_{ab}=\cos\theta_{ab}$, $s_{ab}=\sin\theta_{ab}$, $\theta_{ab}$
are the corresponding vacuum mixing angles, and $\delta_{\mathrm{CP}}$
is the CP violating phase. The values of these parameters can be found
in Ref.~\cite{Sal20}.

The Hamiltonian $H_{1}$ in Eq.~(\ref{eq:Schreq}), which describes
the neutrino interaction with GW, has the form $H_{1}=UH_{m}^{(g)}U^{\dagger}$,
where
\begin{equation}\label{eq:Hgmass}
  H_{m}^{(g)}=H_{m}^{(\mathrm{vac})}\left(A_{c}h_{+}+A_{s}h_{\times}\right),
\end{equation}
is the Hamiltonian in the mass basis, $A_{c}=\tfrac{1}{2}\sin^{2}\vartheta\cos2\varphi\cos[\omega t(1-\cos\vartheta)]$,
$A_{s}=\tfrac{1}{2}\sin^{2}\vartheta\sin2\varphi \sin[\omega t(1-\cos\vartheta)]$,
$h_{+,\times}$ are the amplitudes corresponding to `plus' and `cross'
polarizations of GW, $\omega$ is the frequency of GW, $\vartheta$
and $\varphi$ are the spherical angles fixing the neutrino momentum
with respect to the wave vector of GW, which is supposed to propagate
along the $z$-axis. To derive Eq.~(\ref{eq:Hgmass}) we assume that~\cite{Dvo19}
$\omega L|v_{a}-v_{b}|\ll1$, $a,b=1,2,3$, where $L$ is the distance
of the neutrino beam propagation and $v_{a}$ is the velocity of a
mass eigenstate. Analogously to $H_{m}^{(\mathrm{vac})}$, we subtract
the common diagonal term in $H_{m}^{(g)}$ in Eq.~(\ref{eq:Hgmass}).

The Hamiltonian $H_{m}^{(g)}$ for a circularly polarized GW with
$h_{+}=h_{\times}$ was obtained in Ref.~\cite{Dvo19} based on
the exact solution of the Hamilton-Jacobi equation for a test particle
in a plane GW. In the present work, we provide a more straightforward
perturbative derivation of the same result which is given in Appendix~\ref{sec:DERHG};
cf. Eq.~(\ref{eq:Hgfin}). Of course, the expression for $H_{m}^{(g)}$
coincides with that in Ref.~\cite{Dvo19} in the limit $h_{+}=h_{\times}$.

Now we consider the situation when a neutrino interacts with stochastic
GWs. In this case, the angles $\vartheta$ and $\varphi$, as well
as the amplitudes $h_{+,\times}$, are random functions of time. To
study the neutrino motion in such a background, it is more convenient
to deal with the density matrix $\rho$, which obeys the equation,
$\mathrm{i}\dot{\rho}=[H_{0}+H_{1},\rho]$. Following Ref.~\cite{LorBal94},
we introduce the density matrix in the interaction picture, $\rho_{\mathrm{int}}=\exp(\mathrm{i}H_{0}t)\rho\exp(-\mathrm{i}H_{0}t)$.
It satisfies the equation,
\begin{equation}\label{eq:rhoIeq}
  \mathrm{i}\dot{\rho}_{\mathrm{int}}=
  [H_{\mathrm{int}},\rho_{\mathrm{int}}],
\end{equation}
where $H_{\mathrm{int}}=\exp(\mathrm{i}H_{0}t)H_{1}\exp(-\mathrm{i}H_{0}t)$.
Using the Baker--Campbell--Hausdorff formula and the fact that that
both $H_{m}^{(\mathrm{vac})}$ and $H_{m}^{(g)}$ are diagonal, we
get that $H_{\mathrm{int}}=H_{1}$. This result is valid even before
setting $v_{a}\to1$ in the phase of the wave in Eq.~(\ref{eq:Habint})
and omitting the common factors proportional to the unit matrix in
both $H_{m}^{(\mathrm{vac})}$ and $H_{m}^{(g)}$. The initial condition
for $\rho_{\mathrm{int}}$ coincides with that for $\rho$: $\rho_{\mathrm{int}}(0)=\rho(0)\equiv\rho_{0}$.

We assume that stochastic GWs form a Gaussian random process. In this
situation, all odd correlators of angle factors $A_{c,s}$ and the
amplitudes $h_{+,\times}$ are vanishing. Moreover, we take that $h_{+}$
and $h_{\times}$ are independent. After averaging, the formal solution
of Eq.~(\ref{eq:rhoIeq}) can be present in the form of a series,
\begin{align}\label{eq:serrhoI}
  \left\langle
    \rho_{\mathrm{int}}
  \right\rangle = &
  \rho_{0}-[H_{0},[H_{0},\rho_{0}]]\int_{0}^{t}\mathrm{d}t_{1}\int_{0}^{t_{1}}\mathrm{d}t_{2}
  \big\langle
    \left[
      A_{c}(t_{1})h_{+}(t_{1})+A_{s}(t_{1})h_{\times}(t_{1})
    \right]
    \notag
    \\
    & \times
    \left[
      A_{c}(t_{2})h_{+}(t_{2})+A_{s}(t_{2})h_{\times}(t_{2})
    \right]
  \big\rangle +
  \dotsb
  \nonumber
  \\
  & =
  \rho_{0}-[H_{0},[H_{0},\rho_{0}]]\int_{0}^{t}\mathrm{d}t_{1}\int_{0}^{t_{1}}\mathrm{d}t_{2}
  \big(
    \left\langle
      A_{c}(t_{1})A_{c}(t_{2})
    \right\rangle
    \left\langle
      h_{+}(t_{1})h_{+}(t_{2})
    \right\rangle
    \notag
    \\
    & +
    \left\langle
      A_{s}(t_{1})A_{s}(t_{2})
    \right\rangle
    \left\langle
      h_{\times}(t_{1})h_{\times}(t_{2})
    \right\rangle
  \big)+\dotsb,
\end{align}
where we show only two nonzero terms in order not to encumber the
text.

We can see that the series in Eq.~(\ref{eq:serrhoI}) decays
into two independent ones corresponding to different polarizations
of GW. Each of these series contains only either $\left\langle h_{+}(t)h_{+}(0) \right\rangle$ or $\left\langle h_{\times}(t)h_{\times}(0) \right\rangle$.  In the following, we account for all terms in the expansion in Eq.~\eqref{eq:serrhoI}. Further analysis of each of the series corresponding to different GW polarizations is identical to that in Ref.~\cite{Dvo20}. Therefore we omit the details.

Now, let us consider the averaging of the angle factors. We should mention that both the amplitudes $h_{+,\times}(t)$ and the angles $\vartheta(t)$ and $\varphi(t)$ are random functions of time. Indeed, we consider random distribution of GWs sources. It means that, when a neutrino interacts with a certain GW, the angle between a neutrino momentum and the wave vector of GW is randomly distributed from zero to $\pi$. However, unlike the correlators $\left\langle h_{+,\times}(t_{1})h_{+,\times}(t_{2}) \right\rangle$, which are taken to be arbitrary, we suppose that both $\langle \vartheta(t_1) \vartheta(t_2) \rangle$ and $\langle \varphi(t_1) \varphi(t_2) \rangle$ are proportional to $\delta(t_1 - t_2)$. This supposition is reasonable since it is based on the assumption of the uniform distribution of the sources of GWs in the universe. The form of the correlators $\left\langle h_{+,\times}(t_{1})h_{+,\times}(t_{2}) \right\rangle$ depends on physical processes underlying the GWs production. Thus, it is inexpedient to take that the amplitudes are $\delta$-correlated.

We can study, e.g.,
the correlator $\left\langle A_{c}(t_{1})A_{c}(t_{2})\right\rangle$, which has the form,
\begin{equation}\label{eq:Acav1}
  \left\langle
    A_{c}(t_{1})A_{c}(t_{2})
  \right\rangle =
  \frac{1}{4}
  \left\langle
    \sin^{2}\vartheta_1 \sin^{2}\vartheta_2 \cos(2\varphi_1) \cos(2\varphi_2)
    \cos\alpha_1\cos\alpha_2
  \right\rangle,
\end{equation}
where $\alpha_{1,2} = \omega t_{1,2} (1-\cos\vartheta_{1,2})$ and the angles $\vartheta_{1,2}$ and $\varphi_{1,2}$ correspond to $t_{1,2}$. Since the random variables $\vartheta$ and $\varphi$ are taken to be $\delta$-correlated, we get that $\langle \cos\alpha_1\cos\alpha_2 \rangle = 1/2$. Now, we should average Eq.~\eqref{eq:Acav1} over directions of incoming GWs,
\begin{equation}\label{eq:AcAc}
  \left\langle
    A_{c}(t_{1})A_{c}(t_{2})
  \right\rangle =
  \frac{1}{16\pi^{2}}\int_{0}^{\pi}\mathrm{d}\vartheta\sin^{4}\vartheta \int_{0}^{2\pi}\mathrm{d}\varphi\cos^{2}(2\varphi)=\frac{3}{128}.
\end{equation}
Analogously we show that $\left\langle A_{s}(t_{1})A_{s}(t_{2})\right\rangle = \tfrac{3}{128}$.

The obtained correlators of the angular factors in Eq.~\eqref{eq:AcAc} should be used in Eq.~\eqref{eq:serrhoI}. As we mentioned above, Eq.~\eqref{eq:serrhoI} splits into two independent series. Accounting for all terms in the expansions and applying the results of Ref.~\cite{Dvo20}, we get that
$\left\langle \rho_{\mathrm{int}}\right\rangle $ obeys the equation,
\begin{equation}\label{eq:rhoIeqfin}
  \frac{\mathrm{d}}{\mathrm{d}t}
  \left\langle
    \rho_{\mathrm{int}}
  \right\rangle (t)=
  -g(t)\frac{3}{64}[H_{0},[H_{0},
  \left\langle
    \rho_{\mathrm{int}}
  \right\rangle (t)]],
\end{equation}
where
\begin{equation}\label{eq:gdef}
  g(t)=\frac{1}{2}\int_{0}^{t}\mathrm{d}t_{1}
  \left(
    \left\langle
      h_{+}(t)h_{+}(t_{1})
    \right\rangle +
    \left\langle
      h_{\times}(t)h_{\times}(t_{1})
    \right\rangle
  \right).
\end{equation}
In case of circularly polarized GWs with $h_{+}=h_{\times}$, we reproduce
the results of Ref.~\cite{Dvo20} in Eqs.~(\ref{eq:rhoIeqfin})
and~(\ref{eq:gdef}).

To proceed with the analysis of Eq.~(\ref{eq:rhoIeqfin}), we define
the new matrix $\rho'=U^{\dagger}\left\langle \rho_{\mathrm{int}}\right\rangle U$,
which is the density matrix in the interaction picture for the neutrino
mass eigenstates. After some matrix algebra, we get that $\rho'$
satisfies the equation,
\begin{equation}\label{eq:rho'eq}
  \dot{\rho}'=-\tilde{g}[H_{m}^{(\mathrm{vac})},[H_{m}^{(\mathrm{vac})},\rho']]=-\tilde{g}M,
\end{equation}
where $\tilde{g}=\tfrac{3}{64}g$. The matrix $M$ in Eq.~(\ref{eq:rho'eq})
has the following entries: $M_{ab}=(E_{a}-E_{b})^{2}\rho'_{ab}$. Note
that, here, we use the original form of $H_{m}^{(\mathrm{vac})}$
before the energy decomposition: $\left(H_{m}^{(\mathrm{vac})}\right)_{ab}=E_{a}\delta_{ab}$.

Thus, Eq.~(\ref{eq:rho'eq}) can be integrated straightforwardly
\begin{equation}\label{eq:rho'sol}
  \rho'_{aa}(t) = \rho'_{aa}(0)=\text{const}, \quad
  \rho'_{ab}(t) =\rho'_{ab}(0)g_{ab},\ a\neq b, \quad
  g_{ab}(t) =\exp
  \left[
    -(E_{a}-E_{b})^{2}\int_{0}^{t}\tilde{g}(t')\mathrm{d}t'
  \right],
\end{equation}
where the initial condition $\rho'(0)$ has the form, $\rho'(0)=U^{\dagger}\left\langle \rho_{\mathrm{int}}\right\rangle (0)U=U^{\dagger}\rho(0)U$.
It can be also expressed in the components as $\rho'_{ab}(0)=\sum_{\sigma}U_{\sigma a}^{*}U_{\sigma b}P_{\sigma}(0)$,
where we assume that $\rho_{\lambda\sigma}(0)=\delta_{\lambda\sigma}P_{\sigma}(0)$.
Here the emission probabilities $P_{\sigma}(0)$ are proportional
to the neutrino fluxes at a source: $P_{\sigma}(0)\propto\left(F_{\nu_{\sigma}}\right)_{\mathrm{S}}$.

Accounting for Eq.~(\ref{eq:rho'sol}), we get the expression for
the density matrix for flavor neutrinos, $\rho=UR\rho'R^{\dagger}U^{\dagger}$,
where $\left(R\right)_{ab}=\delta_{ab}\exp(-\mathrm{i}E_{a}t)$. The
probability to detect a certain flavor, after the neutrino beam passes
the distance $x\approx t$, reads $P_{\lambda}(x)=\rho_{\lambda\lambda}(t\approx x)$.
Using the components of $U$ in Eq.~(\ref{eq:U3f}), it can be represented
in the form,
\begin{equation}\label{eq:Plambdag}
  P_{\lambda}^{(g)}(x)=\sum_{\sigma}P_{\sigma}(0)
  \left[
    \sum_{a}|U_{\lambda a}|^{2}|U_{\sigma a}|^{2}+
    2\text{Re}\sum_{a>b}U_{\lambda a}U_{\lambda b}^{*}U_{\sigma a}^{*}U_{\sigma b}\exp
    \left(
      -\mathrm{i}\varphi_{ab}x
    \right)
    g_{ab}
  \right],
\end{equation}
where $\varphi_{ab}=E_{a}-E_{b}=\frac{\Delta m_{ab}^{2}}{2E}$ are
the phases of neutrino vacuum oscillations. Equation~(\ref{eq:Plambdag}),
which takes into account the neutrino interaction with stochastic
GWs, should be compared with the analogous probabilities for neutrino
vacuum oscillations $P_{\lambda}^{(\mathrm{vac})}$, which are derived,
e.g., in Ref.~\cite{GiuKim07}. The difference $\Delta P_{\lambda}=P_{\lambda}^{(g)}-P_{\lambda}^{(\mathrm{vac})}$,
which reveals the effect of GWs on neutrino oscillations, has the
form,
\begin{align}\label{eq:DeltaPgen}
  \Delta P_{\lambda}(x)= & 2\sum_{\sigma}P_{\sigma}(0)
  \sum_{a>b}
  \left\{
    \text{Re}
    \left[
      U_{\lambda a}U_{\lambda b}^{*}U_{\sigma a}^{*}U_{\sigma b}
    \right]
    \cos
    \left(
      2\pi\frac{x}{L_{ab}}
    \right) +
    \text{Im}
    \left[
      U_{\lambda a}U_{\lambda b}^{*}U_{\sigma a}^{*}U_{\sigma b}
    \right]
    \sin
    \left(
      2\pi\frac{x}{L_{ab}}
    \right)
  \right\}
  \nonumber
  \\
  & \times
  \left\{
    1-\exp
    \left[
      -\frac{4\pi^{2}}{L_{ab}^{2}}\int_{0}^{x}\tilde{g}(t)\mathrm{d}t
    \right]
  \right\},
\end{align}
where $L_{ab}=\tfrac{4\pi E}{|\Delta m_{ab}^{2}|}$ are the neutrino
oscillations lengths in vacuum.

If we study the interaction between stochastic GWs and neutrinos emitted
by randomly distributed sources, we have to average Eq.~(\ref{eq:DeltaPgen})
over the propagation distance. It gives $\left\langle \Delta P_{\lambda}\right\rangle =0$.
Therefore, the effect of stochastic GWs on oscillations of such neutrinos
is washed out. The fluxes at a source will coincide with these accounting
for only vacuum oscillations. Thus, the claim in Ref.~\cite{Dvo20},
that stochastic GWs result in small changes of observed fluxes of
neutrinos from randomly distributed sources, is incorrect. The deviation
of fluxes, obtained in Ref.~\cite{Dvo20}, is likely to stem from
an inexactitude of numerical simulations. We avoid this inexactitude
in the present work since we rely on the analytical solution of Eq.~(\ref{eq:rhoIeqfin}).

\section{Application to SN neutrinos\label{sec:APPL}}

In this section, we apply the obtained results for the description
of the propagation of SN neutrinos in the background of stochastic GWs.

A huge amount of energy is carried away
by neutrinos from a core-collapsing SN. The major neutrino luminosity was reported, e.g., in Ref.~\cite{Jan17}
to take place during a $\nu_{e}$-burst, which happens because of
the direct Urca process $e^{-}+p\to n+\nu_{e}$ in the neutronazing
matter of a protoneutron star (PNS). This burst occurs at $\sim(3-4)\,\text{ms}$
after the core bounce and lasts $\lesssim0.1\,\text{s}$~\cite{Jan17}; see also references therein.
The neutrino luminosity can reach $\sim10^{53}\,\text{erg}\cdot\text{s}^{-1}$
during the burst, with almost all of emitted neutrinos being of the
electron type~\cite{Jan17}. 

We start this section with the study of the interaction between stochastic GWs and SN neutrinos emitted in a $\nu_e$-burst. In this situation, the fluxes at
a source are $\left(F_{\nu_{e}}:F_{\nu_{\mu}}:F_{\nu_{\tau}}\right)_{\mathrm{S}}=(1:0:0)$. At later moments of time, other neutrino flavors are emitted. The fluxes of different flavors of SN neutrinos become almost equal by $t \approx (0.05-0.1)\,\text{s}$ after the core bounce. Thus, the initial neutrino fluxes are not in the ratio $(1:0:0)$. The evolution of SN neutrinos with such initial condition in the presence of stochastic GWs is also discussed in this section.

A collapsing star, owing to its relatively small size, can be considered
as an almost point-like neutrino source. Indeed, the size of a neutrinosphere, i.e. an effective sphere where neutrinos are trapped inside,
is $L_{\text{source}}\lesssim100\,\text{km}$ at the moment of a
$\nu_{e}$-burst. The energy of SN neutrinos is $E\sim10\,\text{MeV}$~\cite{VitTamRaf20}.
The oscillations lengths are $L_{21}\approx330\,\text{km}$ and $L_{31}\approx L_{32}\approx10\,\text{km}$
for this energy and $\Delta m_{ab}^{2}$ from Ref.~\cite{Sal20}.
Neutrinos are emitted from any point of a neutrinosphere more or less isotropically. A terrestrial detector can register all SN neutrinos emitted towards it. Thus, we have to integrate the densities of the fluxes over the area of a neutrinosphere, $S_\mathrm{source} \sim L_\mathrm{source}^2$, and divide the result by $S_\mathrm{source}$. We can call this procedure as the averaging over the emission points. Thus,
if we carry out this averaging for $\Delta P_{\lambda}$ in Eq.~(\ref{eq:DeltaPgen}), the (31)- and (32)-contributions are smeared.
The only nonvanishing contribution is from the solar oscillations
channel (21).

As we mentioned above, other neutrino flavors are emitted after a $\nu_e$-burst, changing the
ratio $(1:0:0)$ of the initial fluxes. However, the absolute value of
the SN neutrino luminosity becomes smaller. The neutrinosphere shrinks at these greater times. Nevertheless, its size remains greater than $10\,\text{km}$.
Thus, only the (21)-oscillations channel gives a nonzero contribution
to Eq.~(\ref{eq:DeltaPgen}) even for $t>t_\mathrm{burst}$.

We start by considering neutrinos emitted in a $\nu_e$-burst. Accounting for the ratio of the initial fluxes, we get that $\Delta P_{\lambda}$ for such SN neutrinos takes
the form,
\begin{align}\label{eq:DeltaP21}
  \Delta P_{\lambda}(x)= & 2
  \left[
    \text{Re}
    \left[
      U_{\lambda2}U_{\lambda1}^{*}U_{e2}^{*}U_{e1}
    \right]
    \cos
    \left(
      2\pi\frac{x}{L_{21}}
    \right)+
    \text{Im}
    \left[
      U_{\lambda2}U_{\lambda1}^{*}U_{e2}^{*}U_{e1}
    \right]
    \sin
    \left(
      2\pi\frac{x}{L_{21}}
    \right)
  \right]
  \left[
    1-\exp
    \left(
      -\Gamma
    \right)
  \right],
  \notag
  \\
  \Gamma= & \frac{4\pi^{2}}{L_{21}^{2}}\int_{0}^{x}\tilde{g}(t)\mathrm{d}t.
\end{align}
In Eq.~(\ref{eq:DeltaP21}), we do not set the sine and cosine factors to zero 
despite $x\gg L_{21}$. The propagation distance is huge, but it is fixed.


The correlators of the amplitudes of GWs $\left\langle h_{+,\times}(t)h_{+,\times}(0)\right\rangle $
are related to the spectral density $S(f)$ of GW by~\cite{Chr19}
\begin{equation}\label{eq:S}
  \sum_{ij}
  \left\langle
    h_{ij}(t)h_{ij}(0)
  \right\rangle =
  \left\langle
    h_{+}(t)h_{+}(0)
  \right\rangle +
  \left\langle
    h_{\times}(t)h_{\times}(0)
  \right\rangle =
  \int_{0}^{\infty}\mathrm{d}f\cos(2\pi ft)S(f),
\end{equation}
where $f$ is the frequency measured in Hz. In Eq.~(\ref{eq:S}),
we use Eq.~(\ref{eq:hmunu}). Then, we define the function $\Omega(f)=\tfrac{f}{\rho_{c}}\tfrac{\mathrm{d}\rho_{\mathrm{GW}}}{\mathrm{d}f}$,
where $\rho_{\mathrm{GW}}$ is the energy density of a GW and $\rho_{c}=0.53\times10^{-5}\,\text{Gev}\cdot\text{cm}^{-3}$
is the closure energy density of the universe. Using Eq.~(\ref{eq:S}),
we get that $\Omega(f)=\tfrac{\pi f^{3}}{8\rho_{c}G}S(f)$, where
$G=6.9\times10^{-39}\,\text{GeV}^{-2}$ is the Newton's constant.
The function $\tilde{g}(t)$ has the form,
\begin{equation}\label{eq:tildeg}
  \tilde{g}(t)=\frac{3}{128}\int_{0}^{t}\mathrm{d}t_{1}
  \left(
    \left\langle
      h_{+}(t)h_{+}(t_{1})
    \right\rangle +
    \left\langle
      h_{\times}(t)h_{\times}(t_{1})
    \right\rangle
  \right)=
  \frac{3G\rho_{c}}{32\pi^{2}}\int_{0}^{\infty}\frac{\mathrm{d}f}{f^{4}}\sin(2\pi ft)\Omega(f).
\end{equation}
Now, choosing the source of stochastic GWs, which is fully characterized
by $\Omega(f)$, we can evaluate $\Delta P_{\lambda}$.

We suppose that stochastic GWs are emitted by randomly distributed
merging supermassive BHs. In the case, we can approximate
$\Omega(f)$ by~\cite{Ros11}
\begin{equation}\label{eq:Omegaf}
  \Omega(f)=
  \begin{cases}
    \Omega_{0}, & \text{if}
    \quad
    f_{\mathrm{min}}<f<f_{\mathrm{max}},\\
    0, & \text{otherwise},
  \end{cases}
\end{equation}
where $\Omega_{0}\sim10^{-9}$, $f_{\mathrm{min}}\sim10^{-10}\,\text{Hz}$,
and $f_{\mathrm{max}}\sim10^{-1}\,\text{Hz}$. The main contribution
to $\Gamma$ in Eq.~(\ref{eq:DeltaP21}) results from $f_{\mathrm{min}}$.
Hence we can put $f_{\mathrm{max}}\to\infty$ since $f_{\mathrm{max}}\gg f_{\mathrm{min}}$.

\begin{figure}
  \centering
  \subfigure[]
  {\label{1a}
  \includegraphics[scale=.38]{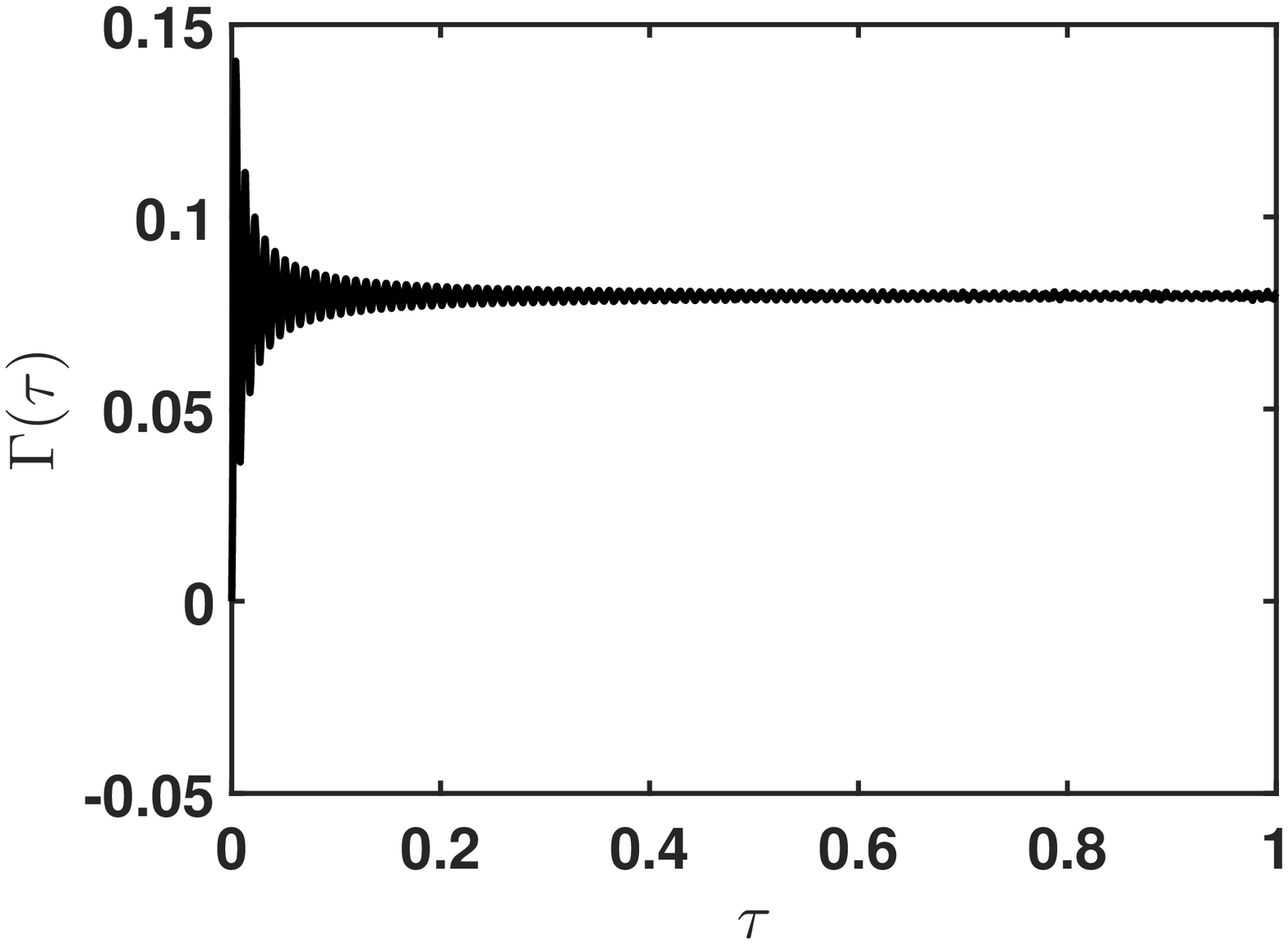}}
  \hskip-.6cm
  \subfigure[]
  {\label{1b}
  \includegraphics[scale=.38]{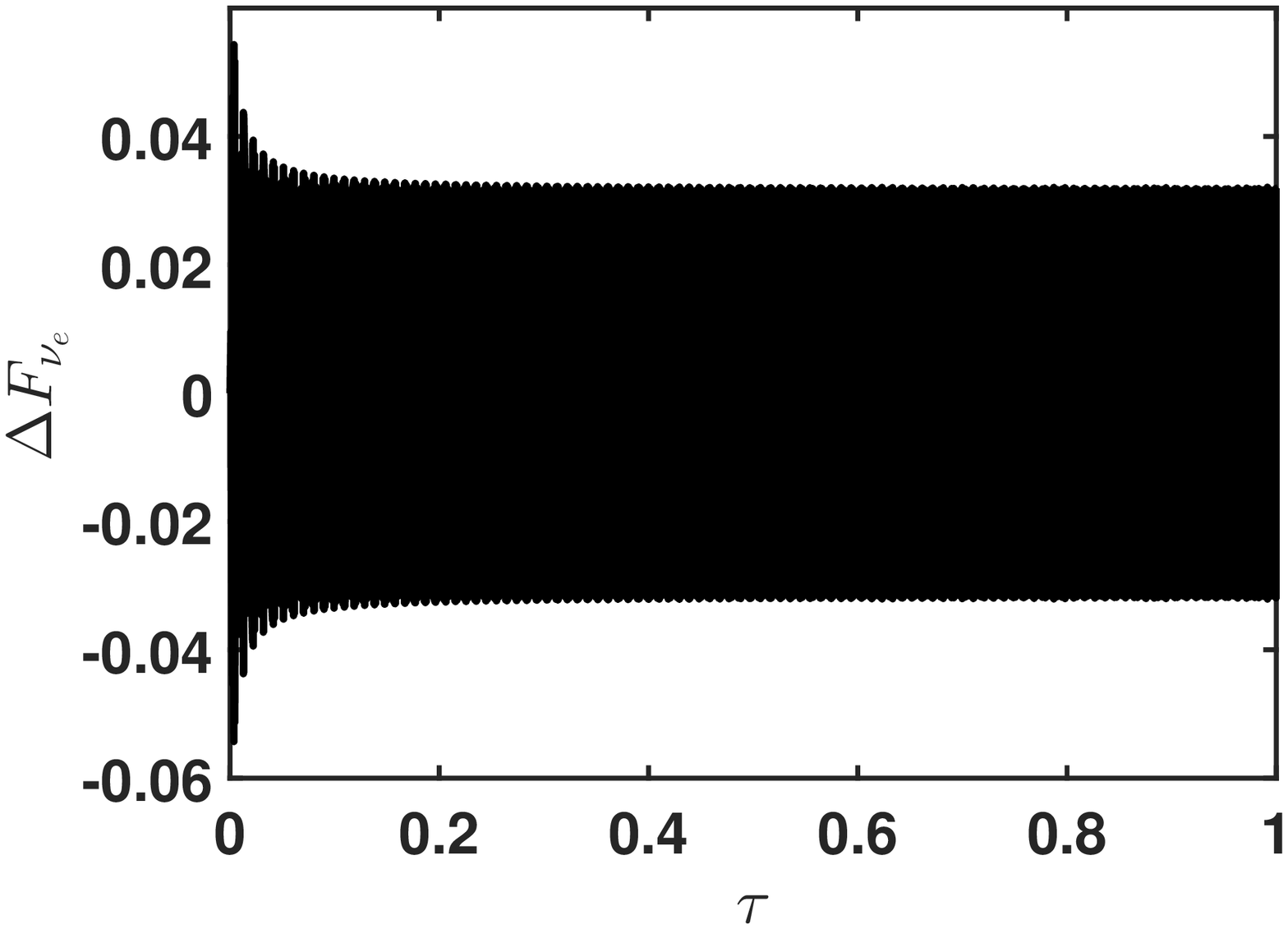}}
  \\
  \subfigure[]
  {\label{1c}
  \includegraphics[scale=.38]{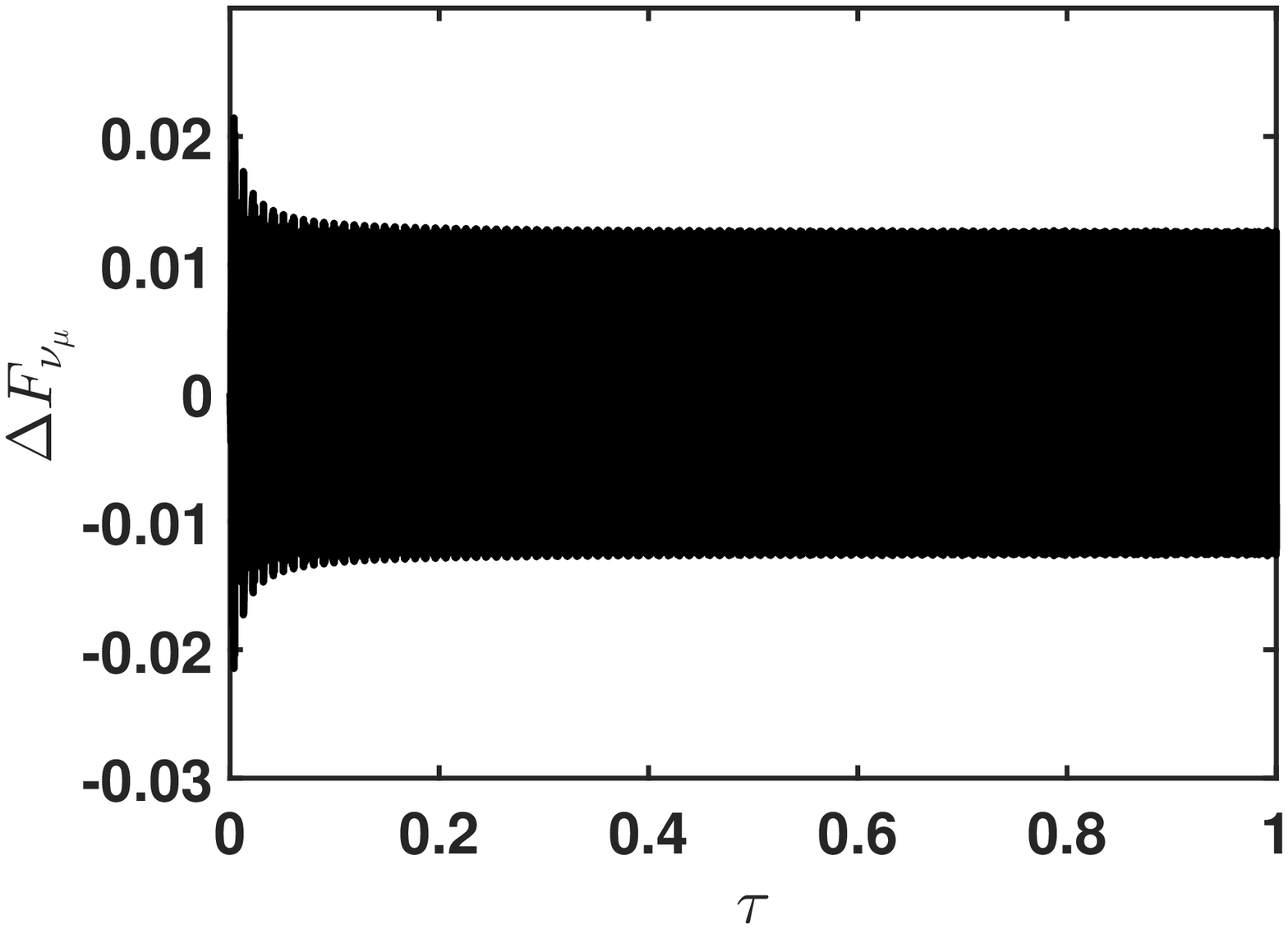}}
  \hskip-.6cm
  \subfigure[]
  {\label{1d}
  \includegraphics[scale=.38]{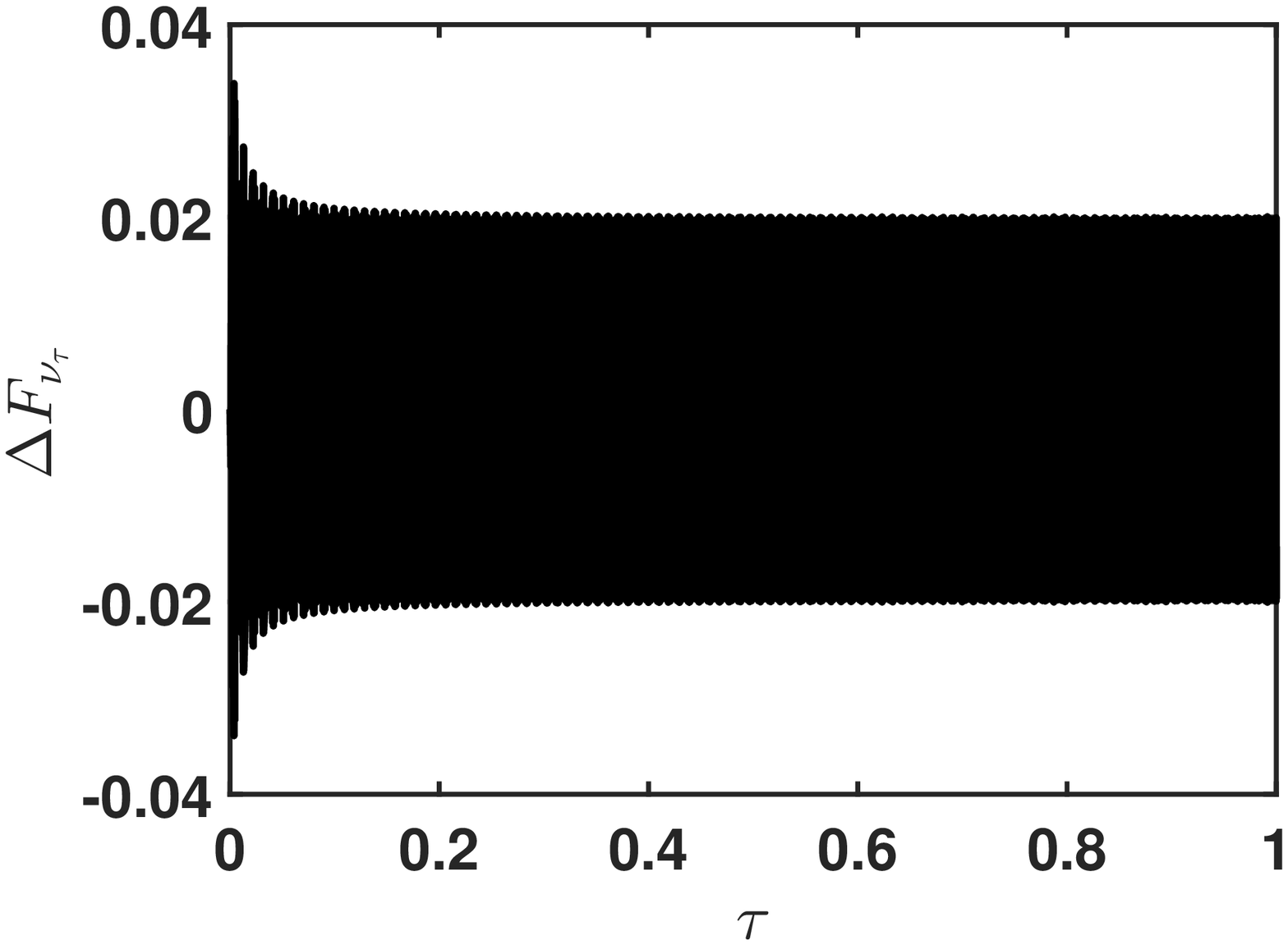}}
  \protect
  \caption{(a) The parameter $\Gamma$ versus the neutrino beam propagation length
  $x=\tau L$; (b)-(d) the corrections the neutrino fluxes $\Delta F_{\nu_{\lambda}}\propto\Delta P_{\lambda}$
  owing to the neutrino interaction with stochastic GWs. The parameters
  of the system are~\cite{VitTamRaf20,Sal20,Ros11} $\Delta m_{21}^{2}=7.5\times10^{-5}\,\text{eV}^{2}$ and
  $E=10\,\text{MeV}$ ($L_{21}=3.3\times10^{2}\,\text{km}$), $\theta_{12}=0.6$,
  $\theta_{23}=0.85$, $\theta_{13}=0.15$, $\delta_{\mathrm{CP}}=3.77$, 
  $L=10\,\text{kpc}$, $\Omega_{0}=10^{-9}$, and $f_{\mathrm{min}}=10^{-10}\,\text{Hz}$.
  The normal neutrino mass ordering is adopted.\label{fig:deltaF}}
\end{figure}

We suppose that $0<x<L$, where $L\sim10\,\text{kpc}$ is the maximal
propagation length, which is taken to be comparable with the Galaxy
size $\sim32\,\text{kpc}$. Using Eqs.~(\ref{eq:DeltaP21}) and~(\ref{eq:tildeg}),
we get the parameter $\Gamma$ in the form,
\begin{equation}\label{eq:Gamma}
  \Gamma = \frac{3G\rho_{c}}{8\pi L_{21}^{2}}
  \int_{f_{\text{min}}}^{f_{\text{max}}}\frac{\mathrm{d}f}{f^{5}}\sin^{2}
  \left(
    \pi fx
  \right)
  \Omega(f)\approx7\times10^{9}\times I
  \left(
    \tau,324
  \right),
  \quad
  I(\tau,\omega_{\text{min}}) = \int_{\omega_{\text{min}}}^{\infty}\frac{\mathrm{d}\omega}{\omega^{5}}\sin^{2}
  \left(
    \omega\tau
  \right)
  \Omega(\omega),
\end{equation}
where $\tau=x/L$ and $\omega_{\text{min}}=\pi Lf_{\text{min}}$ are
the dimensionless parameters. The function $\Gamma(\tau)$ is shown
in Fig.~\ref{1a} for $\Omega(\omega)$ corresponding to
Eq.~(\ref{eq:Omegaf}). In Fig.~\ref{fig:deltaF}, we depict only
the normal mass ordering case since the inverted ordering is almost
excluded experimentally~\cite{Sal18}.

We can see in Fig.~\ref{1a} that $\Gamma$ tends to a
constant value at $\tau\to1$. If we study neutrino fluxes at the
Earth, we put $x=L$, or $\tau=1$. Then, we suppose that the distance
between a source, SN, and a detector, the Earth, is great. It corresponds
to the limit $\omega_{\text{min}}\gg1$. Accounting for Eq.~(\ref{eq:Omegaf}),
we can rewrite $\Gamma\to\Gamma_{\oplus}$ in the form,
\begin{equation}\label{eq:GamEar}
  \Gamma_{\oplus}=\frac{3G\rho_{c}\Omega_{0}}{8\pi f_{\text{min}}^{4}L_{21}^{2}}\omega_{\text{min}}^{4}
  \int_{\omega_{\text{min}}}^{\infty}\frac{\mathrm{d}\omega}{\omega^{5}}\sin^{2}
  \left(
    \omega
  \right) \to
  8\times10^{-2}\times
  \left(
    \frac{\Omega_{0}}{10^{-9}}
  \right)
  \left(
    \frac{f_{\text{min}}}{10^{-10}\,\text{Hz}}
  \right)^{-4}.
\end{equation}
If $\Omega_{0}=10^{-9}$ and $f_{\text{min}}=10^{-10}\,\text{Hz}$,
$\Gamma_{\oplus}=8\times10^{-2}$ in full agreement with Fig.~\ref{1a}.

Although Eq.~(\ref{eq:GamEar}) is valid for $\Omega(f)$ in Eq.~(\ref{eq:Omegaf}),
we can see that $\Gamma_{\oplus}\to0$ at great $f_{\text{min}}$.
This fact explains the result of Ref.~\cite{Dvo20} that stochastic
GWs, emitted by coalescing BHs with stellar masses, do not change
the fluxes of neutrinos. Indeed, in that case~\cite{Ros11}, $f_{\mathrm{min}}\sim10^{-5}\,\text{Hz}\gg10^{-10}\,\text{Hz}$.
Hence $\Gamma_{\oplus}\to0$ and $\Delta P_{\lambda}\to0$ despite
$\Omega$ is greater for such sources.

If $\left(F_{\nu_{e}}:F_{\mu}:F_{\nu_{\tau}}\right)_{\mathrm{S}}=(1:0:0)$
for SN neutrinos emitted in a $\nu_e$-burst, the probabilities at the Earth for vacuum oscillations
are~\cite{GiuKim07}
\begin{equation}\label{eq:Pvac}
  P_{\lambda}^{(\mathrm{vac})}(x)=
  \sum_{a}|U_{\lambda a}|^{2}|U_{ea}|^{2} +
  2\left\{
    \text{Re}
    \left[
      U_{\lambda2}U_{\lambda1}^{*}U_{e2}^{*}U_{e1}
    \right]
    \cos
    \left(
      2\pi\frac{x}{L_{21}}
    \right) +
    \text{Im}
    \left[
      U_{\lambda2}U_{\lambda1}^{*}U_{e2}^{*}U_{e1}
    \right]
    \sin
    \left(
      2\pi\frac{x}{L_{21}}
    \right)
  \right\} ,
\end{equation}
where we accounted for the fact that the contributions of the (31)-
and (32)-oscillations channels are washed out after the averaging over the neutrino emission points on the neutrinosphere surface.
The values of the neutrino fluxes
$F_{\nu_{\lambda}}^{(\mathrm{vac})}\propto P_{\lambda}^{(\mathrm{vac})}$ in Eq.~(\ref{eq:Pvac})
are summarized in Table~\ref{tab:fluxes}. Since the fluxes $F_{\nu_{\lambda}}^{(\mathrm{vac})}$
are rapidly oscillating on the distance $L=10\,\text{kpc}$, in Table~\ref{tab:fluxes},
we present only the mean values and the amplitudes of oscillations:
$(\text{mean value}\pm\text{amplitude})$. We also give $\left(\Delta F_{\nu_{\lambda}}\right)_{\oplus}$
at $x\lesssim L$, i.e. the asymptotic values, which correspond to
Figs.~\ref{1b}-\ref{1d}. The mean value of $\left(\Delta F_{\nu_{\lambda}}\right)_{\oplus}$
equals to zero, as explained above. Thus, we show only the amplitudes
of oscillations of $\left(\Delta F_{\nu_{\lambda}}\right)_{\oplus}$.
One can see in Table~\ref{tab:fluxes} that the relative contribution
of stochastic GWs to the measured neutrino fluxes is at the level
of $(5-7)\,\%$.

\begin{table}
  \centering
  \begin{tabular}{|c|c|c|}
    \hline 
    & $F_{\nu_{\lambda}}^{(\mathrm{vac})}$ & $\left(\Delta F_{\nu_{\lambda}}\right)_{\oplus}$\tabularnewline
    \hline 
    \hline 
    $\nu_{e}$ & $0.5421\pm0.4144$ & $\pm0.0321$\tabularnewline
    \hline 
    $\nu_{\mu}$ & $0.1834\pm0.1636$ & $\pm0.0127$\tabularnewline
    \hline 
    $\nu_{\tau}$ & $0.2745\pm0.2587$ & $\pm0.02$\tabularnewline
    \hline 
  \end{tabular}
  \caption{Second column: the fluxes $F_{\nu_{\lambda}}^{(\mathrm{vac})}$ based
  on Eq.~(\ref{eq:Pvac}) for different neutrino flavors. Third column:
  $\left(\Delta F_{\nu_{\lambda}}\right)_{\oplus}$ corresponding to
  Figs.~\ref{1b}-\ref{1d} for various neutrino types. The parameters
  of neutrinos and GWs are the same as in Fig.~\ref{fig:deltaF}.\label{tab:fluxes}}
\end{table}

Now we turn to the discussion of neutrinos which are emitted after a $\nu_e$-burst, i.e. at $t>(3-4)\,\text{ms}$ after the core bounce. As we mentioned above, the ratio of the emission fluxes is not equal to $(1:0:0)$. At these times, the fluxes of different flavors eventually become almost equal. It happens at $t \approx 0.05\,\text{s}$ after the core bounce (see, e.g., the numerical simulation, carried out in Ref.~\cite{Fis11}). At $t > 0.1\,\text{s}$, the absolute values of the fluxes start to decrease~\cite{Fis11}.

Let us study the influence of stochastic GWs on SN neutrinos emitted at $t_\text{burst} < t < 0.1\,\text{s}$. We can use the general Eq.~\eqref{eq:DeltaPgen} taking that the emission probabilities are time dependent: $P_\sigma(0) \to P_\sigma^{(0)}(\Delta t)$, where $\Delta t = t - t_\text{burst}$ and $t_\text{burst} = (3 - 4)\,\text{ms}$ is the $\nu_e$-burst time. We can approximate $P_\sigma^{(0)}(\Delta t)$ by the following dependence:
\begin{equation}\label{eq:emprob}
  P_{\nu_e}^{(0)}(\Delta t) = \frac{1}{3}
  \left[
    1 + 2 \exp(-5 K\Delta t)
  \right],
  \quad
  P_{\nu_\mu}^{(0)}(\Delta t) = P_{\nu_\tau}^{(0)}(\Delta t) = \frac{1}{3}
  \left[
    1 - \exp(-5 K\Delta t)
  \right],
\end{equation}
where $K = 10\,\text{s}^{-1}$ is the fitting factor. The emission probabilities in Eq.~\eqref{eq:emprob} satisfy $\sum_\sigma P_{\sigma}^{(0)} = 1$ at any $\Delta t$.

\begin{figure}
  \centering
  \subfigure[]
  {\label{2a}
  \includegraphics[scale=.38]{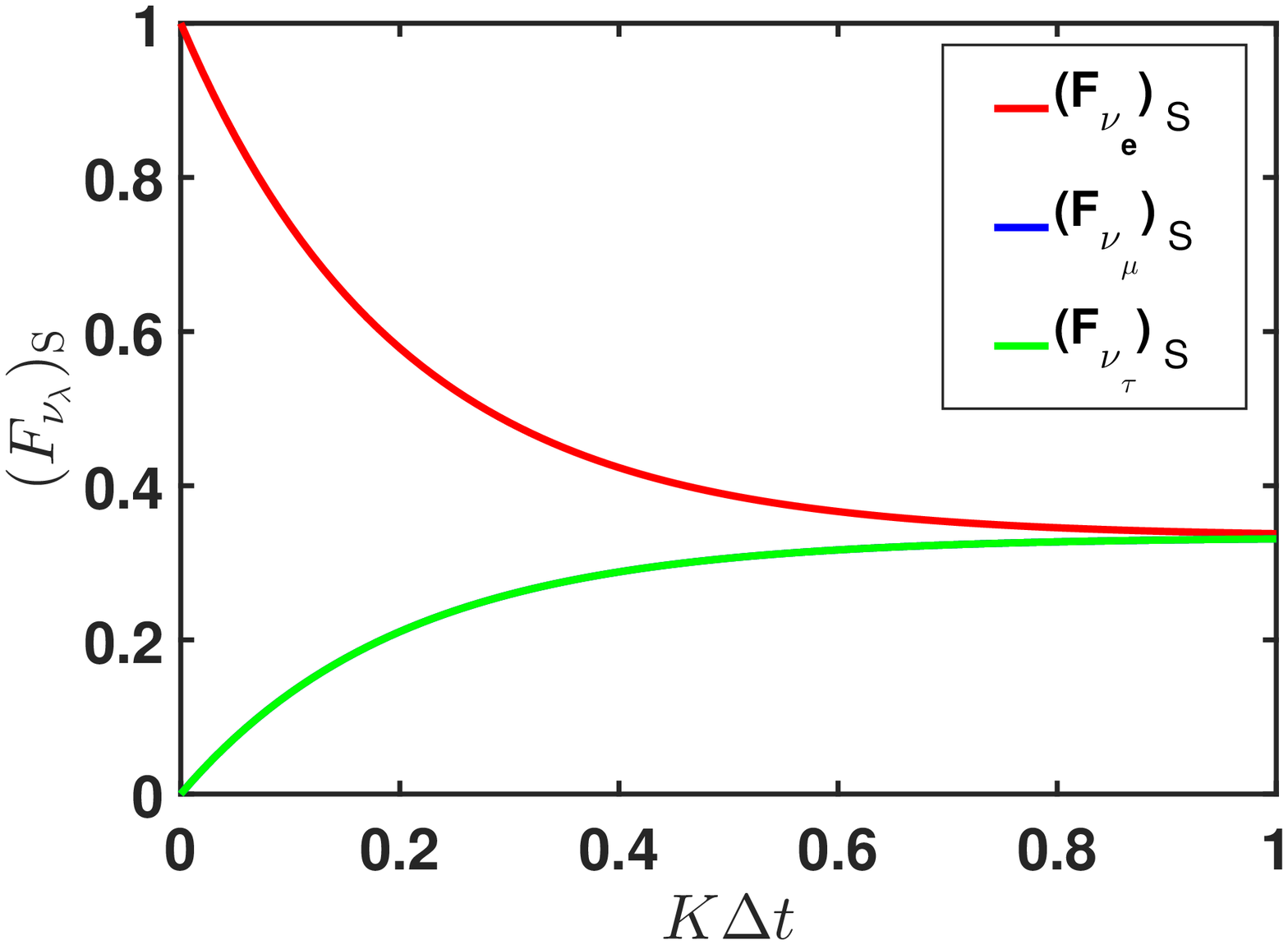}}
  \hskip-.6cm
  \subfigure[]
  {\label{2b}
  \includegraphics[scale=.38]{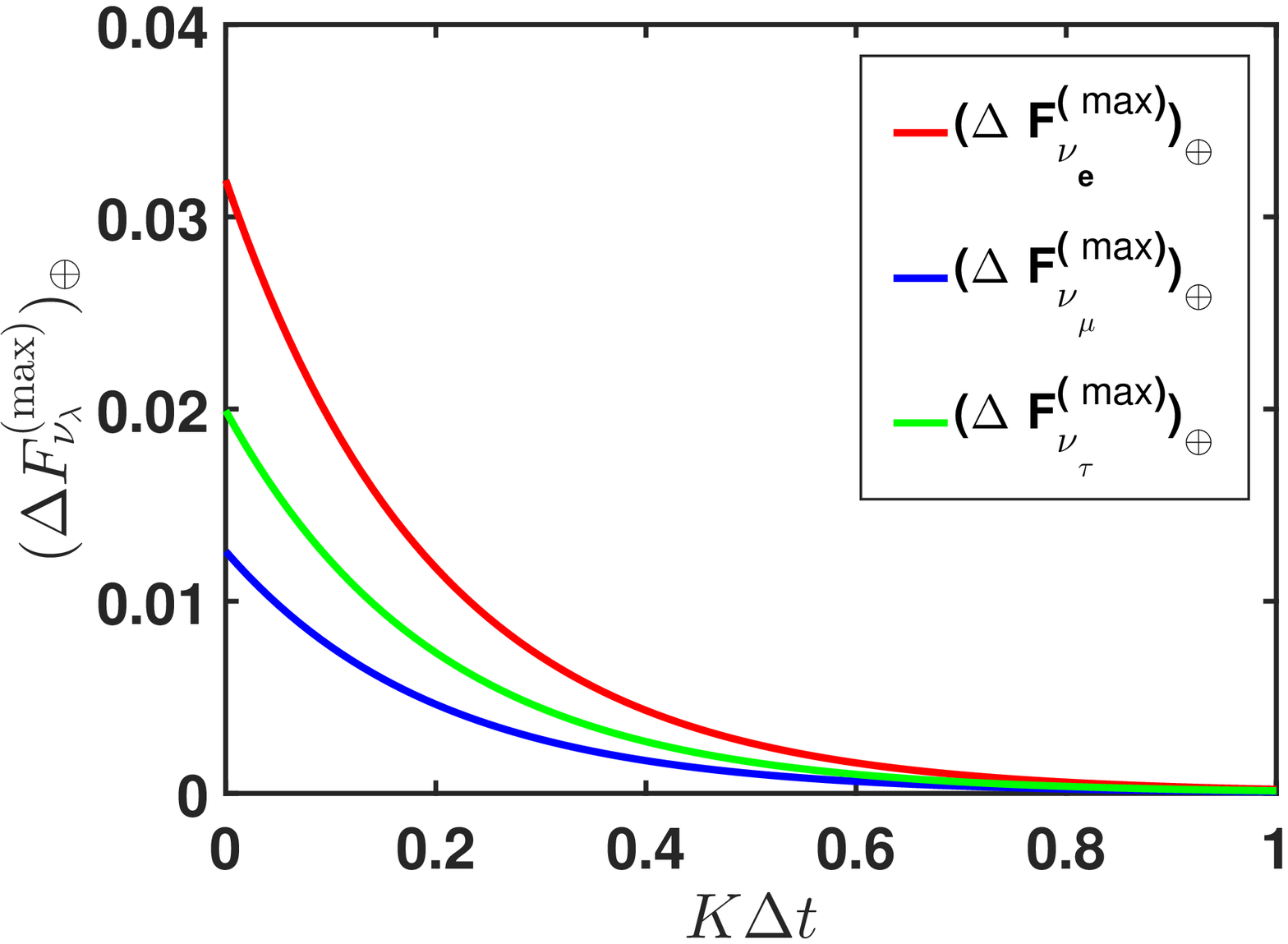}}
  \protect
  \caption{The SN neutrino fluxes for different post bounce times $\Delta t$. The parameters of the neutrino system and GWs are the same as in
  Fig.~\ref{fig:deltaF}; $K\Delta t = 1$ corresponds to $t=0.1\,\text{s}$ after the core bounce.
  (a) The fluxes $(F_{\nu_\lambda})_\text{S}$ at a source. 
  The blue and green lines overlap since the fluxes of $\nu_\mu$ and $\nu_\tau$ coincide [see Eq.~\eqref{eq:emprob}]. 
  (b) The maximal values of the deviations of the fluxes $(\Delta F_{\nu_\lambda}^{(\mathrm{max})})_\oplus$ in a detector, owing to the interaction of SN neutrinos with stochastic GWs.\label{fig:deltaFpb}}
\end{figure}

The initial fluxes $(F_{\nu_\lambda})_\text{S} \propto P_{\nu_\lambda}^{(0)}(\Delta t)$ are shown in Fig~\ref{2a}. The value of $(F_{\nu_e})_\text{S}$ at $\Delta t = 0$ corresponds to the luminosity $\sim 10^{53}\,\text{erg}\cdot\text{s}^{-1}$ in a $\nu_e$-burst.  One can see that $(F_{\nu_\lambda})_\text{S}$ of different flavors become almost equal at $\Delta t \approx K^{-1}$.

The detection probabilities in Eq.~\eqref{eq:DeltaPgen} depend on the propagation length $x$. We consider the maximal values of $\Delta P_{\lambda}$ when neutrinos arrive to a terrestrial detector. They are
\begin{align}\label{eq:DeltaPmax}
  \Delta P_{\lambda}^{(\mathrm{max})}= & 2
    \left[
    1-\exp
    \left(
      -\Gamma_\oplus
    \right)
  \right]
  \sum_{\sigma}P_{\sigma}^{(0)}(\Delta t)
  \sum_{a>b}
  |U_{\lambda a}U_{\lambda b}^{*}U_{\sigma a}^{*}U_{\sigma b}|,
\end{align}
where $\Gamma_\oplus = 8\times 10^{-2}$. Note that $\Delta P_{\lambda}^{(\mathrm{max})}$ in Eq.~\eqref{eq:DeltaPmax} depends on $\Delta t$.

The deviations of the maximal fluxes $(\Delta F_{\nu_\lambda}^{(\mathrm{max})})_\oplus \propto \Delta P_{\lambda}^{(\mathrm{max})}$, owing to the interaction with stochastic GWs, for different post bounce emission times $\Delta t$ are shown in Fig~\ref{2b}. One can see that $(\Delta F_{\nu_\lambda}^{(\mathrm{max})})_\oplus$ at $\Delta t = 0$ coincide with the values given in the third column in Table~\ref{tab:fluxes}. If $\Delta t \to K^{-1}$, $(F_{\nu_\lambda})_\oplus \to 0$. Indeed, $P_\sigma^{(0)} = 1/3$ for any neutrino flavor at these emission times. Thus $\sum_{\sigma}U_{\sigma a}^{*}U_{\sigma b}=\delta_{ab}$ and $\Delta P_{\lambda}^{(\mathrm{max})} = 0$, basing on Eq.~\eqref{eq:DeltaPmax}.

\section{Discussion\label{sec:CONCL}}

In the present work, we have studied the propagation and flavor oscillations
of astrophysical neutrinos interacting with stochastic GWs. We have
rederived more straightforwardly the effective Hamiltonian for such
a system in Appendix~\ref{sec:DERHG}. The analytical expression
for the density matrix of flavor neutrinos has been obtained in Sec.~\ref{sec:DENSMATR}.
Then, in Sec.~\ref{sec:APPL}, we have applied the obtained results
for the description of SN neutrinos.

The present research has several advances compared to Ref.~\cite{Dvo20},
where the interaction of astrophysical neutrinos with stochastic GWs
was also studied. Firstly, we have accounted for two independent polarizations
of GWs contrary to the case of a circularly polarized GW in Ref.~\cite{Dvo20}.
Secondly, now we have found the exact solution of the equation for
the density matrix in the general case of three neutrino flavors. This fact allowed us to correct some statements
about the asymptotic behavior of neutrino fluxes, which were made
basing on numerical simulations in Ref.~\cite{Dvo20}.

In Ref.~\cite{Dvo20}, we studied the interaction between stochastic
GWs and astrophysical neutrinos emitted in decays of charged pions.
Sources of such neutrinos are distributed more or less uniformly in
the universe. Thus one had to average over the neutrino propagation
distance to get the fluxes at the Earth. This fact made it difficult
to extract the contribution of stochastic GWs to neutrino fluxes.
In the present work, we have considered neutrinos emitted in a SN
explosion. It enables us not to perform the averaging over the propagation
distance except for (31)- and (32)-oscillations channels. It is valid
since the oscillations length $L_{21}$ is greater than the radius
of the neutrinosphere, i.e. the size of the neutrino source. The
distance $L$ between a possible SN explosion is great, but it is
fixed. Therefore we should not average over $L/L_{21}$ in the probabilities
for neutrino flavors.

We have also obtained the analytical expression for the damping decrement
$\Gamma$; cf. Eq.~(\ref{eq:GamEar}). This result allowed us to
evaluate the contribution of other sources of stochastic GWs, e.g.,
merging BHs with stellar masses, to the evolution of neutrino fluxes.
The straightforward derivation of the effective Hamiltonian for flavor
neutrinos oscillations, presented in Appendix~\ref{sec:DERHG}, can
be applied to different metrics perturbations besides GWs considered
here. Using this result, we can study, e.g., the evolution of neutrinos
in perturbations in the early universe~\cite{KouMet19,BayPen21}.

In the present work, we have also studied the interaction between stochastic GWs and SN neutrinos emitted in the time interval $t_\mathrm{burst}<t\lesssim0.1\,\text{s}$. When $t \approx 0.1\,\text{s}$, the initial fluxes are almost equal, $\left(F_{\nu_{e}}:F_{\nu_{\mu}}:F_{\nu_{\tau}}\right)_{\mathrm{S}}=(1:1:1)$. We have found in Sec.~\ref{sec:APPL} that, in this situation, the contribution of stochastic GWs to the evolution of SN neutrinos is washed out; cf. Fig.~\ref{2b}. It means that the major effect is for neutrinos emitted at a $\nu_e$-burst, which corresponds to $\left(F_{\nu_{e}}:F_{\nu_{\mu}}:F_{\nu_{\tau}}\right)_{\mathrm{S}}=(1:0:0)$. The neutrino luminosity in a SN explosion remains significant up to $10\,\text{s}$ after the core bounce, which is the time scale for the neutrino driven PNS cooling~\cite{Bur84}. However, the influence of stochastic GWs on SN neutrinos emitted at $t>0.1\,\text{s}$ is vanishing since such neutrinos are emitted with almost equal probabilities. Thus, in Sec.~\ref{sec:APPL}, it is inexpedient to extend $\Delta t$ beyond the $0.1\,\text{s}$ interval.

In Table~\ref{tab:fluxes}, we have summarized the contributions
of stochastic GWs to the fluxes of flavor neutrinos at the Earth.
They are at the level of a few percent. The current neutrino telescopes
are able to detect up to several thousand neutrinos from SN in our
Galaxy~\cite{Sch18}. Future detectors, like the Hyper-Kamiokande,
can detect about $7.5\times10^{4}$ such events~\cite{Abe18}. Thus,
the interaction with stochastic GWs can results in the change of the
SN neutrinos fluxes by $\sim\pm350$ events, in case of the Super-Kamiokande,
and by $\sim\pm3750$ events, for the Hyper-Kamiokande.


\begin{acknowledgments}
I am thankful to A.~V.~Yudin and J.~W.~F.~Valle for the communications, as well as to V.~A.~Berezin for the useful discussion. The work
is supported by the government assignment of IZMIRAN.
\end{acknowledgments}

\appendix

\section{Derivation of the effective Hamiltonian\label{sec:DERHG}}

The action $S_{a}(\mathbf{x},t)$ of a neutrino mass eigenstate with
the mass $m_{a}$, moving in the curved spacetime with the metric
$g_{\mu\nu}$, obeys the Hamilton-Jacobi equation,
\begin{equation}\label{eq:HJeq}
  g_{\mu\nu}\frac{\partial S_{a}}{\partial x_{\mu}}\frac{\partial S_{a}}{\partial x_{\nu}}=m_{a}^{2}.
\end{equation}
The metric in Eq.~(\ref{eq:HJeq}) is supposed to have the small
perturbation $h_{\mu\nu}$ of the Minkowski flat metric $\eta_{\mu\nu}=\text{diag}(+1,-1,-1,-1)$:
$g_{\mu\nu}=\eta_{\mu\nu}+h_{\mu\nu}$.

We look for the solution of Eq.~(\ref{eq:HJeq}) in the form, $S_{a}=S_{a}^{(0)}+S_{a}^{(1)}+\dotsc.$
Here, the zero order term $S_{a}^{(0)}$ obeys the equation, $\eta_{\mu\nu}\partial^{\mu}S_{a}^{(0)}\partial^{\nu}S_{a}^{(0)}=m_{a}^{2}.$
The finction $S_{a}^{(0)}$ has the form, $S_{a}^{(0)}(x)=p_{a}^{\mu}x_{\mu}$, where
$p_{a}^{\mu}=(E_{a},\mathbf{p})$, $E_{a}=\sqrt{m_{a}^{2}+\mathbf{p}^{2}}$,
and $\mathbf{p}$ is the constant neutrino momentum.

In the linear approximation in $h_{\mu\nu}$, the first order term
$S_{a}^{(1)}$ obeys the equation,
\begin{equation}\label{eq:2ord}
  2\eta_{\mu\nu}p_{a}^{\mu}\frac{\partial S_{a}^{(1)}}{\partial x_{\nu}}+h_{\mu\nu}p_{a}^{\mu}p_{a}^{\nu}=0.
\end{equation}
We can transform the term $\eta_{\mu\nu}p_{a}^{\mu}\partial^{\nu}S_{a}^{(1)}$
in Eq.~(\ref{eq:2ord}) as
\begin{equation}\label{eq:alongtr}
  \eta_{\mu\nu}p_{a}^{\mu}\frac{\partial S_{a}^{(1)}}{\partial x_{\nu}}=E_{a}
  \left(
    \frac{\partial S_{a}^{(1)}}{\partial t}+
    \left(
      \mathbf{v}_{a}\nabla
    \right)
    S_{a}^{(1)}
  \right)=
  E_{a}\frac{\mathrm{d}S_{a}^{(1)}}{\mathrm{d}t}.
\end{equation}
where we use $\mathbf{v}_{a}\equiv\mathbf{v}_{a}^{(0)}=\mathbf{p}/E_{a}$
as the particle velocity. It means that a neutrino is supposed to
move along a straight line. In general situation, one has that $\mathbf{v}_{a}=\mathbf{v}_{a}^{(0)}+\delta\mathbf{v}_{a}(t)$
in curved spacetime. However, since $\delta\mathbf{v}_{a}\propto h_{\mu\nu}$,
this term can be neglected in Eq.~(\ref{eq:alongtr}) because $S_{a}^{(1)}\propto h_{\mu\nu}$ already.

Finally, we get that $\frac{\mathrm{d}S_{a}^{(1)}}{\mathrm{d}t}=-\frac{1}{2E_{a}}h_{\mu\nu}p_{a}^{\mu}p_{a}^{\nu}$.
The contribution to the effective Hamiltonian can be obtained as $\left(H_{m}\right)_{aa}=\tfrac{\mathrm{d}S_{a}}{\mathrm{d}t}$.
One can check the validity of this expression in the vaccum case.
Thus, we obtain that the neutrino intraction with a gravitational
field, induced by a metric perturbation, contributes to the effective
Hamiltonian in the mass basis as
\begin{equation}\label{eq:Haa}
  \left(H_{m}^{(g)}\right)_{ab}=-\frac{\delta_{ab}}{2E_{a}}h_{\mu\nu}p_{a}^{\mu}p_{a}^{\nu}.
\end{equation}
Note that we have to take $h_{\mu\nu}$ on the particle trajectory,
$h_{\mu\nu}(\mathbf{x},t)=h_{\mu\nu}(\mathbf{x}(t),t)$.

If we choose a plane GW as the metric perturbation, one has that~\cite{Buo07}
\begin{equation}\label{eq:hmunu}
  h_{ij}=
  \left(
    \begin{array}{cc}
      h_{+}\cos\phi_{a} & h_{\times}\sin\phi_{a}\\
      h_{\times}\sin\phi_{a} & -h_{+}\cos\phi_{a}
    \end{array}
  \right),
  \quad
  i,j=1,2,
\end{equation}
and $h_{0\mu}=h_{3\mu}=0$. Here $\phi_{a}=\omega t-kz=\omega t(1-v_{a}\cos\vartheta)$
is the phase of the wave accounting for the neutrino motion and $h_{+,\times}$
are the amplitudes of different polarizations of GW. In Eq.~(\ref{eq:hmunu}),
we take the Cartesian coordinates $x^{\mu}=(t,\mathbf{x})$. Using
Eqs.~(\ref{eq:Haa}) and~(\ref{eq:hmunu}), we obtain that
\begin{align}\label{eq:Habint}
  \left(
    H_{m}^{(g)}
  \right)_{ab}= &
  -\frac{p^{2}}{2E_{a}}\delta_{ab}\sin^{2}\vartheta
  \left[
    h_{+}\cos2\varphi\cos\phi_{a}+h_{\times}\sin2\varphi\sin\phi_{a}
  \right],
  \nonumber
  \\
  & \to\frac{m_{a}^{2}}{2E}\frac{\delta_{ab}}{2}\sin^{2}\vartheta
  \left[
    h_{+}\cos2\varphi\cos\phi+h_{\times}\sin2\varphi\sin\phi
  \right],
\end{align}
where $\phi=\omega t(1-\cos\vartheta)$. To derive Eq.~(\ref{eq:Habint})
we assume that $\omega L|v_{a}-v_{b}|\ll1$~\cite{Dvo19} and omit
the common factor propotional to the unit matrix. Finally, one gets
that
\begin{equation}\label{eq:Hgfin}
  H_{m}^{(g)}=H_{m}^{(\mathrm{vac})}\left(A_{c}h_{+}+A_{s}h_{\times}\right),
\end{equation}
where $A_{c}=\tfrac{1}{2}\sin^{2}\vartheta\cos2\varphi\cos[\omega t(1-\cos\vartheta)]$
and $A_{s}=\tfrac{1}{2}\sin^{2}\vartheta\sin2\varphi\sin[\omega t(1-\cos\vartheta)]$.


\begin{thebibliography}{50}

\bibitem{Nobel2015}
  T.~Kajita,
  Nobel Lecture: Discovery of atmospheric neutrino oscillations,
  Rev. Mod. Phys. \textbf{88}, 030501 (2016);
  A.~B.~McDonald, Nobel Lecture: The Sudbury Neutrino Observatory: Observation of flavor change for solar neutrinos,
  ibid. \textbf{88}, 030502 (2016).

\bibitem{Smi05}
  A.~Yu.~Smirnov,
  The MSW effect and Matter Effects in Neutrino Oscillations,
  Phys. Scr. \textbf{T121}, 57--64 (2005)
  [hep-ph/0412391].

\bibitem{Giu19}
  C.~Giunti, K.~A.~Kouzakov, Y.-F.~Li, A.~V.~Lokhov, A.~I.~Studenikin, and S.~Zhou,
  Electromagnetic neutrinos in laboratory experiments and astrophysics,
  Ann. Phys. (Amsterdam) \textbf{528}, 198--215 (2016)
  [arXiv:1506.05387].

\bibitem{PirRoyWud96}
  D.~P\'{\i}riz, M.~Roy, and J.~Wudka,
  Neutrino Oscillations in Strong Gravitational Fields,
  Phys. Rev. D \textbf{54}, 1587--1599 (1996)
  [hep-ph/9604403].

\bibitem{CarFul97}
  C.~Y.~Cardall and G.~M.~Fuller,
  Neutrino oscillations in curved space-time: An heuristic treatment,
  Phys. Rev. D \textbf{55}, 7960--7966 (1997)
  [hep-ph/9610494].

\bibitem{For97}
  N.~Fornengo, C.~Giunti, C.~W.~Kim, and J.~Song,
  Gravitational effects on the neutrino oscillation,
  Phys. Rev. D \textbf{56}, 1895--1902 (1997)
  [hep-ph/9611231].

\bibitem{Qua16}
  J.~Q.~Quach,
  Spin gravitational resonance and graviton detection,
  Phys. Rev. D \textbf{93}, 104048 (2016)
  [arXiv:1602.03837].

\bibitem{ObuSilTer17}
  Yu.~N.~Obukhov, A.~J.~Silenko, and O.~V.~Teryaev,
  General treatment of quantum and classical spinning particles in external fields,
  Phys. Rev. D \textbf{96}, 105005 (2017)
  [arXiv:1708.05601].

\bibitem{Dvo19a}
  M.~Dvornikov,
  Neutrino spin oscillations in external fields in curved spacetime,
  Phys. Rev. D \textbf{99}, 116021 (2019)
  [arXiv:1902.11285].

\bibitem{KouMet19}
  G.~Koutsoumbas and D.~Metaxas,
  Neutrino oscillations in gravitational and cosmological backgrounds,
  Gen. Relativ. Gravit. \textbf{52}, 102 (2020)
  [arXiv:1909.02735].

\bibitem{Dvo19}
  M.~Dvornikov,
  Neutrino flavor oscillations in stochastic gravitational waves,
  Phys. Rev. D \textbf{100}, 096014 (2019)
  [arXiv:1906.06167].

\bibitem{Dvo20}
  M.~Dvornikov,
  Flavor ratios of astrophysical neutrinos interacting with stochastic gravitational waves having arbitrary spectra,
  J. Cosmol. Astropart. Phys. \textbf{12} (2020) 022
  [arXiv:2009.02195].

\bibitem{Sch18}
  K.~Scholberg,
  Supernova signatures of neutrino mass ordering,
  J. Phys. G: Nucl. Part. Phys. \textbf{45}, 014002 (2018)
  [arXiv:1707.06384].

\bibitem{Abb16}
  LIGO Scientific Collaboration and Virgo Collaboration,
  B.~P.~Abbott et al.,
  Observation of gravitational waves from a binary black hole merger,
  Phys. Rev. Lett. \textbf{116}, 061102 (2016)
  [arXiv:1602.03837].

\bibitem{Alb19}
  ANTARES, IceCube, LIGO, Virgo Collaborations,
  A.~Albert et al.,
  Search for Multi-messenger Sources of Gravitational Waves and High-energy Neutrinos with Advanced LIGO during its first Observing Run,
  ANTARES and IceCube,
  Astrophys. J. \textbf{870}, 134 (2019)
  [arXiv:1810.10693].

\bibitem{Aar20a}
  IceCube Collaboration,
  M.~G.~Aartsen et al.,
  IceCube Search for Neutrinos Coincident with Compact Binary Mergers from LIGO-Virgo's First Gravitational-Wave Transient Catalog,
  Astrophys. J. Lett. \textbf{898}, L10 (2020)
  [arXiv:2004.02910].

\bibitem{Aar20b}
  IceCube-Gen2 Collaboration,
  M.~G.~Aartsen et al.,
  IceCube-Gen2: The Window to the Extreme Universe,
  J. Phys. G: Nucl. Part. Phys. \textbf{48}, 060501 (2021)
  [arXiv:2008.04323].

\bibitem{Per19}
  B.~B.~P.~Perera et al.,
  The International Pulsar Timing Array: Second data release,
  Mon. Not. R. Astron. Soc. \textbf{490}, 4666--4687 (2019)
  [arXiv:1909.04534].

\bibitem{Arz20}
  NANOGrav Collaboration,
  Z.~Arzoumanian et al.,
  The NANOGrav 12.5-year Data Set: Search For An Isotropic Stochastic Gravitational-Wave Background,
  Astrophys. J. Lett. \textbf{905}, L34 (2020)
  [arXiv:2009.04496].

\bibitem{RomCor17}
  J.~D.~Romano and N.~J.~Cornish,
  Detection methods for stochastic gravitational-wave backgrounds: A unified treatment,
  Living Relativity \textbf{20}, 2 (2017)
  [arXiv:1608.06889].

\bibitem{Raf96}
  G.~G. Raffelt,
  \textit{Stars as Laboratories for Fundamental Physics: The Astrophysics of Neutrinos, Axions, and Other Weakly Interacting Particles}
  (Chicago, University of Chicago Press, 1996),
  pp.~414--430.

\bibitem{VitTamRaf20}
  E.~Vitagliano, I.~Tamborra, and G.~Raffelt,
  Grand unified neutrino spectrum at Earth: Sources and spectral components,
  Rev. Mod. Phys. \textbf{92}, 045006 (2020)
  [arXiv:1910.11878].

\bibitem{An16}
  F.~An et al.,
  Neutrino physics with JUNO,
  J. Phys. G: Nucl. Part. Phys. \textbf{43}, 030401 (2016)
  [arXiv:1507.05613].

\bibitem{Sal20}
  P.~F.~de~Salas, D.~V.~Forero, S.~Gariazzo, P.~Martnez-Mirave, O.~Mena, C.~A.~Ternes, M.~Tortola, and J.~W.~F.~Valle,
  2020 Global reassessment of the neutrino oscillation picture,
  J. High Energy Phys. \textbf{02} (2021) 071
  [arXiv:2006.11237].

\bibitem{LorBal94}
  F.~N.~Loreti and A.~B.~Balantekin,
  Neutrino oscillations in noisy media,
  Phys. Rev. D \textbf{50}, 4762--4770 (1994)
  [nucl-th/9406003].

\bibitem{GiuKim07}
  C.~Giunti and C.~W.~Kim,
  \textit{Fundamentals of Neutrino Physics and Astrophysics}
  (Oxford, Oxford University Press, 2007),
  pp.~247--252.

\bibitem{Jan17}
  H.-Th.~Janka,
  Neutrino Emission from Supernovae,
  in \textit{Handbook of Supernovae}, ed. by A.~W.~Alsabti and P.~Murdin
  (Cham, Springer, 2017),
  pp.~1575--1604
  [arXiv:1702.08713].


\bibitem{Chr19}
  N.~Christensen,
  Stochastic gravitational wave backgrounds,
  Rep. Prog. Phys. \textbf{82}, 016903 (2019)
  [arXiv:1811.08797].

\bibitem{Ros11}
  P.~A.~Rosado,
  Gravitational wave background from binary systems,
  Phys. Rev. D \textbf{84}, 084004 (2011)
  [arXiv:1106.5795].

\bibitem{Sal18}
  P.~F.~de~Salas, S.~Gariazzo, O.~Mena, C.~A.~Ternes, and M.~Tortola,
  Neutrino Mass Ordering from Oscillations and Beyond: 2018 Status and Future Prospects,
  Front. Astron. Space Sci. \textbf{5}, 36 (2018)
  [arXiv:1806.11051].

\bibitem{Fis11}
  T.~Fischer, G.~Mart{\'{\i}}nez-Pinedo, M.~Hempel, and M.~Liebend\"{o}rfer,
  Neutrino spectra evolution during proto-neutron star deleptonization,
  Phys. Rev. D \textbf{85}, 083003 (2012)
  [arXiv:1112.3842].

\bibitem{BayPen21}
  G.~Baym and J.-C.~Peng,
  Evolution of Primordial Neutrino Helicities in Cosmic Gravitational Inhomogeneities,
  Phys. Rev. D \textbf{103}, 123019 (2021)
  [arXiv:2103.11209].

\bibitem{Bur84}
  A.~Burrows,
  On detecting stellar collapse with neutrinos,
  Astrophys. J. \textbf{283}, 848--852 (1984).

\bibitem{Abe18}
  Hyper-Kamiokande Proto-Collaboration,
  K.~Abe et al.,
  Physics potentials with the second Hyper-Kamiokande detector in Korea,
  Prog. Theor. Exp. Phys. \textbf{2018}, 063C01 (2018)
  [arXiv:1611.06118].

\bibitem{Buo07}
  P.~Hoyng,
  \textit{Relativistic Astrophysics and Cosmology: A Primer}
  (Berlin, Springer, 2006),
  pp.~133--136.
  
\end{thebibliography}
\end{document}